\documentclass[final,1p,times,authoryear]{elsarticle}

\usepackage{graphicx}
\usepackage{amsmath, amssymb, amsthm}
\usepackage{natbib}
\usepackage{url}
\journal{Astronomy $\&$ Computing}

\begin{document}

\begin{frontmatter}



\title{Self-Supervised ConvLSTM for Fermi Large Area Telescope Transient Detection\tnoteref{t1,t2}}

\tnotetext[t1]{\textcopyright{} 2026. This manuscript version is made available under the CC BY-NC-ND 4.0 license \url{https://creativecommons.org/licenses/by-nc-nd/4.0/}.}

\tnotetext[t2]{This is the accepted manuscript version of an article published in \emph{Astronomy and Computing}, vol.~56, p.~101128, 2026. The version of record is available at \url{https://doi.org/10.1016/j.ascom.2026.101128}. Please cite the published version: Garinei et al., \emph{Astronomy and Computing}, 56, 101128 (2026), \url{https://doi.org/10.1016/j.ascom.2026.101128}.}


\author[idea,unimarconi]{Alberto Garinei}

\author[idea]{Stefano Speziali\corref{cor1}}
\ead{sspeziali@idea-re.eu}
\cortext[cor1]{Corresponding author.}

\author[idea]{Alessandro Vispa}

\author[idea]{Andrea Marini}

\author[unipg]{Sara Cutini}

\author[idea]{Emanuele Piccioni}

\author[idea,unimarconi]{Marcello Marconi}

\author[unitrieste,uniudine]{Francesco Longo}

\author[unimarconi]{Matteo Martini}

\author[unimarconi]{Francesca Fallucchi}

\author[unimarconi]{Romeo Giuliano}

\author[unimarconi]{Ernesto William De Luca}

\author[unimarconi]{Umberto Di Matteo}

\author[unimarconi]{Sabino Meola}

\affiliation[idea]{organization={Idea-Re S.r.l., 06125},
            city={Perugia},
            country={Italy}}

\affiliation[unimarconi]{organization={Dipartimento Scienze Ingegneristiche, Università Guglielmo Marconi, 00193},
            city={Rome},
            country={Italy}}
            
\affiliation[unipg]{organization={Istituto Nazionale Fisica Nucleare, sezione di Perugia, 06123},
            city={Perugia},
            country={Italy}}

\affiliation[unitrieste]{organization={Dipartimento di Fisica, Università di Trieste, I-34127},
            city={Trieste},
            country={Italy}}

\affiliation[uniudine]{organization={Università di Udine and INFN Trieste, I-33100},
            city={Udine},
            country={Italy}}

\begin{abstract}
We present a framework for detecting transient $\gamma$-ray phenomena in a controlled environment by combining end-to-end simulations of the Fermi-LAT sky with self-supervised spatio-temporal deep learning. We generate a ten-year synthetic Universe with \texttt{gtobssim} and process the simulated events into daily all-sky maps of counts and exposure, obtaining a time-ordered sequence that mirrors the structure of Fermi-LAT observations. To model the nominal evolution of the sky, we employ a Convolutional Long Short-Term Memory (ConvLSTM) network that operates directly on map sequences, preserving spatial locality while learning temporal dependencies. The model is trained to reconstruct expected emission, and departures from the learned baseline are quantified through pixel-wise mean-squared residual maps. We then define statistically motivated anomaly criteria by estimating per-pixel thresholds from the residual distribution on the training set, and we enforce spatial coherence via local filtering to suppress isolated fluctuations. The ConvLSTM  is then deployed as trained predictor on Fermi-LAT daily maps, where the sky can depart from the nominal behavior because of genuine astrophysical variability and instrumental non-stationarities. The resulting pipeline flags localized, time-dependent excesses consistent with high-variable sources or transient events (e.g., flares or GRBs) and provides a benchmark for evaluating anomaly-detection strategies on long-duration, Fermi-LAT-like datasets.
\end{abstract}



\begin{keyword}
Transient sources \sep Gamma-ray astronomy \sep ConvLSTM \sep Anomaly detection \sep Spatio-temporal analysis



\end{keyword}

\end{frontmatter}




\section{Introduction}
\label{sec:introduction}

The detection and characterization of transient gamma--ray activity is a central challenge in high--energy astrophysics. Localized flares, bursts, and episodes of rapid variability can develop on timescales ranging from hours to days, and their signatures are often superimposed on structured diffuse emission and on a photon--limited, energy--dependent background. In this regime, robust transient identification requires methods that can separate genuinely significant deviations from fluctuations driven by limited counting statistics, exposure variations, and background mismodeling.

The \textit{Fermi} Large Area Telescope (Fermi-LAT) has enabled continuous monitoring of the gamma--ray sky since 2008, supporting repeated all-sky coverage and systematic variability studies on short cadences \citep{Atwood2009LAT,ackermann2013fava}. This observational capability, combined with the volume and diversity of the accumulated dataset, motivates automated, data-driven approaches for identifying statistically significant departures from nominal emission.

In many realistic settings, fully supervised classification is hindered by the scarcity of comprehensive labels for transient phenomena, while the very definition of ``anomaly'' is intrinsically contextual, depending on timescale, sky location, and expected background conditions \citep{Chandola2009Anomaly}. This has stimulated broad interest in unsupervised and weakly supervised anomaly detection, including reconstruction--based methods that learn typical patterns and flag departures from them \citep{Ruff2021ADReview}.

In this work, we adopt a controlled strategy to study transient detection by building a long-term synthetic gamma--ray sky observed with Fermi-LAT-like survey geometry. Starting from \texttt{gtobssim} simulations within the Fermitools ecosystem \citep{ascl:1905.011}, and interfacing higher-level analysis workflows when relevant through community tools such as \texttt{fermipy} \citep{Wood2018Fermipy}, we produce daily full-sky maps of counts and exposure and use them to train convolutional recurrent neural networks, focusing on ConvLSTM architectures \citep{Shi2015ConvLSTM}. The networks are trained to reconstruct sequences of daily sky maps, effectively learning the nominal spatio-temporal evolution of the simulated universe. Reconstruction residuals between predicted and true frames---quantified via pixel--wise mean--squared errors---are then interpreted as anomaly scores, designed to be sensitive to localized excess emission.

Building on this reconstruction--based perspective, we introduce a statistically grounded anomaly--detection pipeline. Residual distributions estimated on the training set are used to derive adaptive decision thresholds on a pixel-wise basis, intended to reflect the expected reconstruction error under nominal conditions. Candidate anomalies in the test set are identified as statistically unlikely excursions above these thresholds; spatial coherence criteria are then applied to suppress isolated noise fluctuations and emphasize clustered residual structures.

Because the ConvLSTM network is trained on \texttt{gtobssim} simulations to learn an average nominal spatio-temporal baseline of the gamma--ray sky, we finally apply the trained model to Fermi-LAT daily maps, where the observed emission may deviate from the learned baseline due to both genuine astrophysical variability and instrumental or operational non-stationarities.

The main contributions of this study are:
\begin{itemize}
    \item the construction of a realistic, long-term synthetic dataset in Fermi-LAT observation format for reproducible anomaly-detection experiments;
    \item the use of ConvLSTM networks to learn the nominal spatiotemporal behavior of daily gamma--ray sky maps via sequence reconstruction;
    \item a residual--based anomaly detection framework combining statistically motivated thresholding with spatial coherence filtering;
    \item an evaluation of the method's ability to recover localized transient phenomena within a fully controlled simulation environment.
\end{itemize}

The remainder of this paper is organized as follows. Section~\ref{sec:fermi_data_tools} introduces the Fermi-LAT data products and the simulation pipeline adopted in this work. Section~\ref{sec:methodology} details the neural-network methodology used to carry out anomaly detection. Section~\ref{sec:results} discusses the performance and limitations of the method. Finally, Section~\ref{sec:conclusion} outlines conclusions and directions for future research.

\section{Fermi-LAT Data and Simulation Pipeline}
\label{sec:fermi_data_tools}

\subsection{Fermi-LAT data products and analysis context}
\label{sec:fermi_data_context}

Since its launch in 2008, the \textit{Fermi} Gamma-ray Space Telescope has provided long-term monitoring of the high-energy sky. Its primary instrument, the Large Area Telescope (LAT), is a wide field--of--view pair--conversion telescope sensitive from below $\sim 20$~MeV to beyond $\sim 300$~GeV, designed to support long-term surveys, source catalogs, and time-domain studies \citep{Atwood2009LAT}.
Over the mission lifetime, the publicly distributed Fermi-LAT event reconstruction and classification have undergone major upgrades---most notably the Pass~8 event-level analysis---which improved effective area, background rejection, and angular/energy reconstruction \citep{Atwood2013Pass8,Bruel2018Pass8}. These developments underpin the modern Fermi-LAT data products and catalogs, including the 3FGL and 4FGL generations and their incremental updates \citep{Acero2015_3FGL,Abdollahi2020_4FGL,Abdollahi2022_4FGLDR3}. At the same time, dedicated efforts such as the Fermi All--sky Variability Analysis (FAVA) illustrate the value of automated variability searches at the all-sky level \citep{ackermann2013fava}.

From the user perspective, Fermi-LAT science analyses typically start from a small set of standardized data products distributed by the \textit{Fermi} Science Support Center (FSSC). In particular, Fermi-LAT observations are provided as FITS files \citep{Pence2010FITS} whose structure is tailored to support reproducible filtering, exposure computation, and likelihood-based modeling. The core Fermi-LAT products relevant to this work can be summarized as follows:

\begin{description}
    \item[Event (FT1) files.]
    Event files (often referred to as FT1) store reconstructed photon candidates and their per-event attributes, including arrival time, reconstructed energy, and reconstructed direction. They also include analysis-relevant classifications (e.g., event class and event type) and quality-related metadata that enable standardized selections and downstream modeling \citep{ascl:1905.011}. Because the event reconstruction evolves across data releases, the interpretation of these fields is tied to the adopted Pass and IRF set \citep{Atwood2013Pass8,Bruel2018Pass8}.

    \item[Spacecraft (FT2) files.]
    Spacecraft files (FT2) describe the observatory state as a function of time, including position and attitude, as well as quantities used to derive livetime and good-time intervals. These products are essential to translate event counts into exposure-corrected measurements and to account for periods of reduced data quality (e.g., passages through the South Atlantic Anomaly) \citep{ascl:1905.011}. In practice, consistent temporal coverage between FT1 and FT2 is required for reliable exposure calculations and time selections.

    \item[Instrument Response Functions (IRFs).]
    IRFs provide the mapping between incident photons and reconstructed events, typically factorized into effective area, point spread function (PSF), and energy dispersion \citep{FermiLATPerformance}. The IRFs are defined for each event class and event type partition, reflecting analysis choices that trade acceptance against residual background and reconstruction quality. Throughout this work we adopt the Pass~8 Release~3 IRF family (P8R3\_V3), consistent with the SOURCE-class analysis configuration commonly used for survey and catalog-style studies.

    \item[Source catalogs.]
    Fermi-LAT catalogs provide curated lists of detected sources with positions, spectra, and variability characterizations, serving both as astrophysical references and as practical inputs for model construction. In this work we refer to the third and fourth Fermi-LAT catalog generations (3FGL and 4FGL) and to the incremental 4FGL data releases that extend the temporal baseline and refine modeling \citep{Acero2015_3FGL,Abdollahi2020_4FGL,Abdollahi2022_4FGLDR3, Ballet2023_4FGLDR4}.
\end{description}

Taken together, these datasets enable analyses that range from all-sky population studies to targeted investigations of individual sources and transient activity. Importantly, the separation between photon events (FT1), pointing/livetime information (FT2), and response modeling (IRFs) makes it possible to propagate instrumental effects transparently and to reproduce selections and exposures across independent studies \citep{FermiLATPerformance}.

\subsection{Fermi tools and \texttt{fermipy}}
\label{sec:fermi_tools}

The Fermi mission provides a dedicated software suite---historically referred to as the \textit{Fermi Science Tools} and currently distributed as the \texttt{Fermitools}---to perform standard Fermi-LAT analyses, including event filtering, map binning, exposure calculation, and likelihood fitting \citep{ascl:1905.011}. The FSSC maintains extensive online documentation and analysis threads describing recommended workflows and best practices for common analysis tasks. Higher-level analysis workflows can be interfaced through community-developed tools such as \texttt{fermipy} \citep{Wood2018Fermipy}.

A typical Fermi-LAT workflow begins with the definition of a valid event sample through spatial, temporal, and quality selections (e.g., energy range, zenith-angle cuts, event class/type, and good-time intervals). These steps are implemented through command-line tools that operate directly on FITS products and are designed to compose into reproducible pipelines \citep{ascl:1905.011}. Higher-level products such as binned counts maps, exposure cubes, and source maps can then be generated and used in model fitting.

Model fitting in Fermi-LAT analyses is commonly carried out within a maximum-likelihood framework, where the predicted sky model (sources plus diffuse components, folded with exposure and IRFs) is compared to the observed data; likelihood-ratio based test statistics provide a standard measure of source significance and model comparison \citep{Mattox1996Likelihood}. While our work focuses on map-based learning and residual-based detection, the same conceptual ingredients---data selection, IRFs, and exposure modeling---underpin both classical likelihood analyses and data-driven approaches.

\subsection{\texttt{gtobssim} simulation setup}
\label{sec:gtobssim_setup}

To develop and validate transient-search strategies in a controlled environment, we generated a long-duration synthetic dataset in Fermi-LAT format with \texttt{gtobssim}, the Monte Carlo observation simulator distributed within the Fermi Science Tools (Fermitools) \citep{ascl:1905.011}. In \texttt{gtobssim}, photons are sampled from a user-defined sky model---specified through an XML description that may include point-like and extended sources as well as diffuse components---and are then propagated through the instrument response. For this, we contributed to the \texttt{fermimodel} Python package\footnote{The main function of this package is to produce model files for use in analysis with gtlike and simulations with gtobssim. See \url{https://fermi.gsfc.nasa.gov/ssc/data/analysis/user/} and \url{https://github.com/ideare-ds/fermimodel}.} for use with the Fermitools. Instrumental effects are incorporated via the chosen Instrument Response Functions (IRFs), which encode the effective area, point-spread function, and energy dispersion, thereby producing event lists with realistic energy and directional uncertainties \citep{FermiLATPerformance}.

A key input to \texttt{gtobssim} is the observing history (spacecraft pointing and livetime), which determines the time-dependent exposure pattern. In this work we anchored the simulation timeline to the beginning of the mission's nominal survey operations, following the completion of the initial on-orbit checkout on 2008-08-04. We simulated 3649 daily time bins, i.e. approximately ten years of survey observations, consistent with the cadence adopted by the downstream pipeline.

We adopted the IRFs \texttt{P8R3\_SOURCE\_V3} throughout the analysis \citep{FermiLATPerformance}. The input XML model included the standard diffuse components recommended for Pass~8 SOURCE-class analyses, namely the Galactic interstellar emission template \texttt{gll\_iem\_v07.fits} and the isotropic template \texttt{iso\_P8R3\_SOURCE\_V3\_v1.txt} \citep{FSSCBackgroundModels}, together with a population of catalogued sources defining our synthetic sky.

\begin{figure}[h!!]
    \centering
    \includegraphics[width=0.7\linewidth]{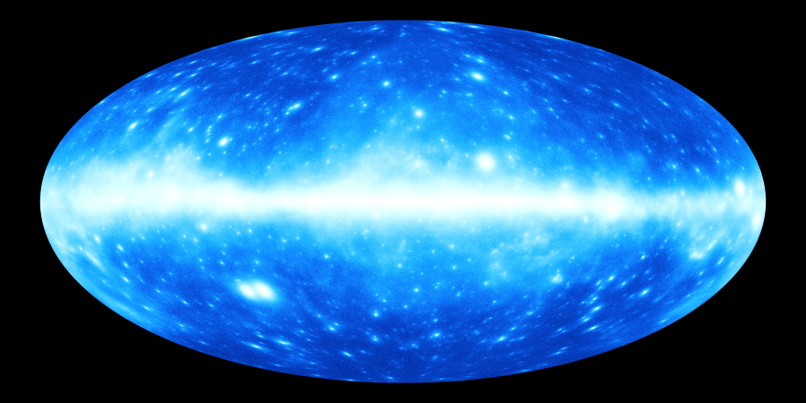}
    \caption{All-sky counts map integrated over the full simulated period ($\sim$10 years) in Galactic coordinates (Mollweide projection). The bright band traces the Galactic diffuse emission, while localized enhancements correspond to simulated point sources.}
    \label{fig:fullsky}
\end{figure}

A practical limitation is that a monolithic, all-sky simulation over $\sim$10 years can be prohibitively demanding in memory and wall time, since the computational cost scales with both the number of model components and the simulated duration. To make production tractable, we adopted a tiling and chunking strategy. The celestial sphere was partitioned into a set of 192 sky tiles. For sky tiling we adopt the equal-area HEALPix discretization \citep{Gorski2005HEALPix}. Each tile was simulated using an XML model that included all sources within the tile plus an angular buffer region (overlap) to mitigate boundary effects associated with the finite PSF. The full temporal interval was further split into consecutive segments of duration \(\Delta t \simeq 90\)~days, such that each \texttt{gtobssim} job simulated one tile over one temporal chunk. These jobs were executed in parallel across multiple CPU cores. In a final aggregation step, events were geometrically filtered to the nominal (non-overlapping) tile footprints and then merged across tiles and time chunks to reconstruct a continuous, ten-year, all-sky event dataset suitable for standard Fermi-LAT binning and exposure calculations.
In Figure~\ref{fig:fullsky} we show our full sky simulation using the \texttt{gtobssim} tool.

\subsection{Generated dataset}
\label{sec:generated_dataset}

The output of our simulation and binning pipeline is a five-dimensional tensor of shape
\[
(3649,\; 2,\; 10,\; 360,\; 180),
\]
representing a daily synthetic dataset of the gamma-ray sky as observed by the Fermi-LAT over \(\sim\)10 years\footnote{The dataset can be downloaded at \url{https://huggingface.co/datasets/Idea-re/fermi-lat-synthetic-daily-sky-maps}.}. The first axis (\(3649\)) corresponds to one full-sky frame per day. The second axis contains the two map products used as input channels in the learning task: counts and exposure. The third axis (\(10\)) corresponds to logarithmically spaced energy bins covering 100~MeV--500~GeV, while the last two axes (\(360\times180\)) encode an all-sky map in celestial coordinates with a spatial resolution of one degree per pixel. Of course, a more refined spatial resolution is possible and, in some sense, needed at the expense of more expensive calculations. We discuss this further at the end of the paper.

Count maps were produced with \texttt{gtbin} using the \texttt{CCUBE} algorithm \citep{ascl:1905.011}, where we adopted a Cartesian (CAR) projection in celestial coordinates with a $360\times180$ pixel grid (1$^\circ$/pixel).

Exposure maps were computed on the same spatial and spectral grid by first generating livetime cubes with \texttt{gtltcube} and then running \texttt{gtexpcube2} \citep{ascl:1905.011}. We used \texttt{coordsys=CEL} and the IRFs \texttt{P8R3\_SOURCE\_V3} \citep{FermiLATPerformance}. This ensures that exposure is fully consistent with the response assumptions adopted during event generation and that the two channels can be combined coherently in downstream analyses. The simulated event lists and the resulting binned products are written in FITS format, mirroring standard Fermi-LAT analysis inputs.

To keep this first study focused and computationally tractable, we also consider an energy-integrated version of the dataset obtained by summing over the energy axis. This produces a tensor of shape
\[
(3649,\; 2,\; 360,\; 180),
\]
which is the representation used in the remainder of the paper. The motivation for this choice is twofold. First, it allows us to isolate the spatio-temporal anomaly-detection problem and establish a baseline end-to-end pipeline without introducing the additional complexity of modeling spectral variability across multiple sparse energy bins. Second, it reduces the dimensionality of the learning task, making training and model selection more manageable in this initial study. We stress, however, that this simplification does not alter the general ConvLSTM framework adopted here; rather, it defines the scope of the present work. Extending the method to energy-resolved inputs is a natural and important next step, since spectral information may further improve sensitivity to classes of transients characterized by energy-dependent variability.

\section{Neural Network Methodology}
\label{sec:methodology}

\subsection{ConvLSTM for spatio-temporal sky modeling and sequence construction}
\label{sec:convlstm_modeling}

To model the spatio-temporal evolution of daily gamma-ray sky maps we adopt
Convolutional Long Short-Term Memory (ConvLSTM) networks \citep{Shi2015ConvLSTM},
a recurrent architecture that replaces the fully-connected transformations of
standard LSTMs \citep{Hochreiter1997LSTM} with convolutional operators. This
choice preserves the 2D structure of the maps while learning temporal
dependencies, which is particularly suitable for transient searches where
anomalies appear as localized, time-dependent excesses.

In the following, we use ConvLSTM in a self-supervised \emph{next-frame
prediction} setting: given a short history of consecutive daily maps, the model
predicts the subsequent map. Transient activity not explained by the learned
nominal dynamics is expected to produce localized prediction residuals, which
we later convert into anomaly scores.

Let the full time series of maps be represented as
\[
  \mathbf{X} \in \mathbb{R}^{N \times C \times H \times W},
\]
where $N$ is the number of daily frames, $C$ the number of channels, and $H,W$
the spatial dimensions. For a chosen window length $\texttt{seq\_len}$, we build
overlapping sequences, yielding
\[
  N_{\mathrm{win}} = N - \texttt{seq\_len} + 1
\]
training windows. Each window $i=0,\dots,N_{\mathrm{win}}-1$ is split into
\begin{align*}
  \mathbf{x}^{(i)} &\in \mathbb{R}^{(\texttt{seq\_len}-1)\times C\times H\times W}
  &&\text{(input frames)}\\
  \mathbf{y}^{(i)} &\in \mathbb{R}^{C\times H\times W}
  &&\text{(target frame)}
\end{align*}
with
\[
  \mathbf{x}^{(i)} = \bigl[\mathbf{X}[i],\,\dots,\,\mathbf{X}[i+\texttt{seq\_len}-2]\bigr],
  \quad
  \mathbf{y}^{(i)} = \mathbf{X}[i+\texttt{seq\_len}-1].
\]
The ConvLSTM receives $\mathbf{x}^{(i)}$ and outputs a prediction
$\hat{\mathbf{y}}^{(i)}$ for the next frame. Of course, in our case $C = 2$, $(H, W) = (180, 360)$, whereas \texttt{seq\_len} has been chosen to be equal to 5. We set \texttt{seq\_len}=5 as a practical compromise between temporal context and computational cost. This choice provides a multi-day context for prediction while keeping memory usage and training time within a manageable range for the present study.

\subsection{ConvLSTM architecture and hyperparameter search}
\label{sec:convlstm_arch_hpo}

The model consists of $L$ ConvLSTM layers with $C_h$ hidden channels and
convolutional kernel size $k \times k$. The final hidden representation is
mapped to the output space through a lightweight convolutional head, implemented
as a sequence of $1\times1$ or $3\times3$ convolutions with channel widths
specified by \texttt{final\_conv\_channels}, ending with $C$ output channels.
The architectural hyperparameters explored are therefore
\[
\{C_h,\, k,\, L,\, \texttt{final\_conv\_channels}\}.
\]

We performed a grid search over the learning rate and a discrete set of ConvLSTM
configurations. Specifically, we tested
\[
\texttt{learning\_rate} \in \{10^{-3},\,10^{-4},\,10^{-5}\},
\]
and the following five architecture variants:
\begin{itemize}
    \item \textbf{Config A:} \texttt{input\_channels}=2, \texttt{hidden\_channels}=32, \texttt{kernel\_size}=3, \texttt{num\_layers}=1, \texttt{final\_conv\_channels}=[16, 2].
    \item \textbf{Config B:} \texttt{input\_channels}=2, \texttt{hidden\_channels}=32, \texttt{kernel\_size}=3, \texttt{num\_layers}=2, \texttt{final\_conv\_channels}=[16, 2].
    \item \textbf{Config C:} \texttt{input\_channels}=2, \texttt{hidden\_channels}=64, \texttt{kernel\_size}=3, \texttt{num\_layers}=1, \texttt{final\_conv\_channels}=[32, 2].
    \item \textbf{Config D:} \texttt{input\_channels}=2, \texttt{hidden\_channels}=64, \texttt{kernel\_size}=5, \texttt{num\_layers}=2, \texttt{final\_conv\_channels}=[32, 16, 2].
    \item \textbf{Config E:} \texttt{input\_channels}=2, \texttt{hidden\_channels}=32, \texttt{kernel\_size}=5, \texttt{num\_layers}=3, \texttt{final\_conv\_channels}=[16, 8, 2].
\end{itemize}
For each combination of architecture and learning rate, we trained the network
on the training split and selected the best-performing model on the validation
set.

The optimal setup across the explored grid was obtained with
\[
\texttt{learning\_rate}=10^{-3}
\]
and \textbf{Config C}:
\begin{center}
\texttt{\{input\_channels: 2,\ hidden\_channels: 64,\ kernel\_size: 3,\ num\_layers: 1,\ final\_conv\_channels: [32, 2]\}}.
\end{center}
Model selection was based on validation performance. We defined the best-performing setup as the one achieving the most favorable validation-loss behavior while maintaining a reasonable computational cost. Under this criterion, \texttt{learning\_rate}$=10^{-3}$ was retained for the subsequent comparison of model configurations, while Config~C was selected as the most balanced solution in terms of reconstruction quality, model complexity, and training time, as discussed in Section~\ref{sec:results}.

\subsection{Threshold estimation and anomaly detection}
\label{sec:threshold}

Given an input window of length $\texttt{seq\_len}$, the model predicts the next
frame $\hat{\mathbf{y}}^{(i)}\in\mathbb{R}^{C\times H\times W}$, which is compared
to the true target $\mathbf{y}^{(i)}$. We compute the pixel-wise squared error
\[
  e^{(i)}_{c,h,w}
  = \bigl(\hat y^{(i)}_{c,h,w} - y^{(i)}_{c,h,w}\bigr)^2,
  \quad
  c=1,\dots,C,\;h=1,\dots,H,\;w=1,\dots,W.
\]
A 2D residual map is then obtained by averaging over channels in the usual manner:
\[
  m^{(i)}_{h,w}
  = \frac{1}{C}\sum_{c=1}^C e^{(i)}_{c,h,w}.
\]

Residual statistics are heterogeneous across the sky because both count statistics
and exposure vary spatially. For this reason, anomaly thresholds are estimated
separately for each pixel using training residuals only. Let $T_{\mathrm{train}}$
denote the number of frames assigned to the training set, and let \texttt{seq\_len}
be the window length. Since the dataset is split into training and validation subsets
with an 80:20 ratio, the number of residual maps available for threshold calibration is
\[
  N_{\mathrm{train,win}} = T_{\mathrm{train}} - \texttt{seq\_len} + 1.
\]
Collecting the training residual maps $\{m^{(i)}\}_{i=1}^{N_{\mathrm{train,win}}}$
yields, for each pixel $(h,w)$, an empirical distribution
\[
  \mathcal{E}(h,w)=\bigl\{m^{(i)}_{h,w}\mid i=1,\dots,N_{\mathrm{train,win}}\bigr\}.
\]
We define a per-pixel threshold map as the $p$-th empirical quantile:
\[
T_{\mathrm{perc}}(h,w)
\;=\;
\mathrm{percentile}\!\bigl(\mathcal{E}(h,w),\,p\bigr),
\qquad p=95\%.
\]
We set $p=95\%$ as a conservative per-pixel quantile calibrated on nominal training residuals, which fixes the expected pixel-wise exceedance rate to $5\%$ under the null; the subsequent spatial-coherence filter then suppresses isolated exceedances and controls the overall candidate rate. At inference time, a pixel is flagged as anomalous whenever
\[
m_{h,w} \;>\; T_{\mathrm{perc}}(h,w),
\]
yielding a binary anomaly mask
\[
  M_{h,w}
  =
  \begin{cases}
    1, & m_{h,w} > T_{\mathrm{perc}}(h,w),\\[2pt]
    0, & \text{otherwise}.
  \end{cases}
\]

To suppress isolated fluctuations and emphasise spatially coherent structures,
we optionally apply a local density filter to the mask: for each pixel, a window
of size $W\times W = 8 \times 8$ is examined, and the pixel is retained as anomalous only if
at least $K = 4$ pixels within the window are above threshold. We set the spatial-coherence filter to a $W \times W = 8 \times 8$ neighborhood with $K=4$ in order to suppress isolated pixel-level excursions while retaining compact multi-pixel structures: requiring at least four suprathreshold pixels within an $8 \times 8$ vicinity provides a lightweight morphological prior that favors spatially coherent residual patterns without being overly restrictive on their shape.
This enforces spatial coherence, ensuring that only compact ``blobs'' of elevated residuals survive. If the filtered mask contains one or more anomalous pixels (or connected components), the corresponding frame is labeled as anomalous and regions are
highlighted for inspection.

\section{Results and Discussion}
\label{sec:results}

In this section we report the performance of the ConvLSTM model selected through the grid search described in Section~\ref{sec:convlstm_arch_hpo}, namely \textbf{Config C} with \(\texttt{learning\_rate}=10^{-3}\). The network is trained in a self-supervised setting to predict the next daily all-sky frame from the preceding \(\texttt{seq\_len}-1\) frames, minimizing a pixel-wise mean squared error (MSE). After training, the prediction residuals provide the anomaly score maps employed by the statistical detection procedures introduced in Section~\ref{sec:threshold}.

\subsection{Neural Network Performance}
\label{sec:nn_performance}

\begin{figure}[htbp]
    \centering
    \begin{minipage}[t]{0.49\linewidth}
        \centering
        \includegraphics[width=\linewidth]{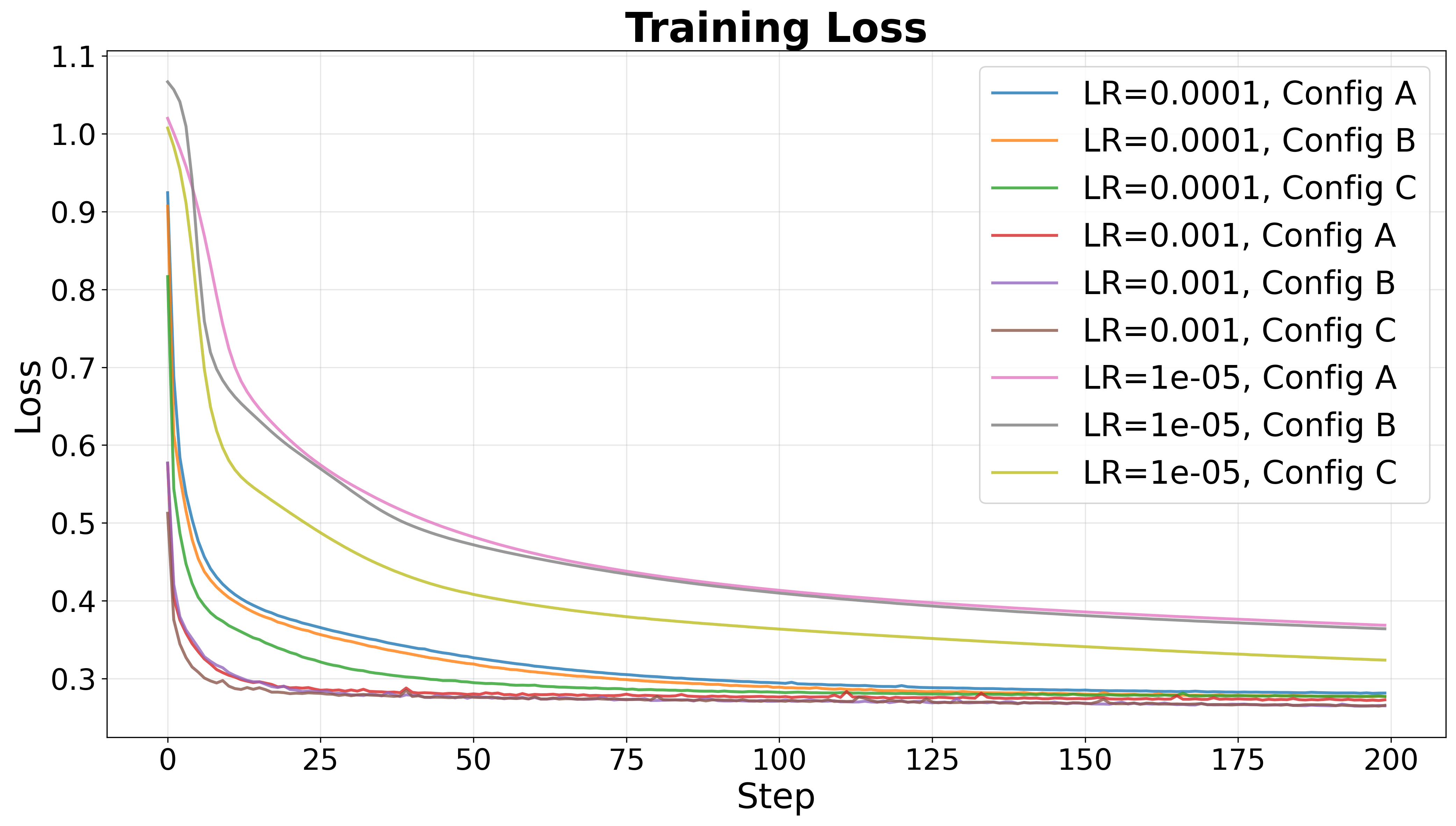}
    \end{minipage}\hfill
    \begin{minipage}[t]{0.49\linewidth}
        \centering
        \includegraphics[width=\linewidth]{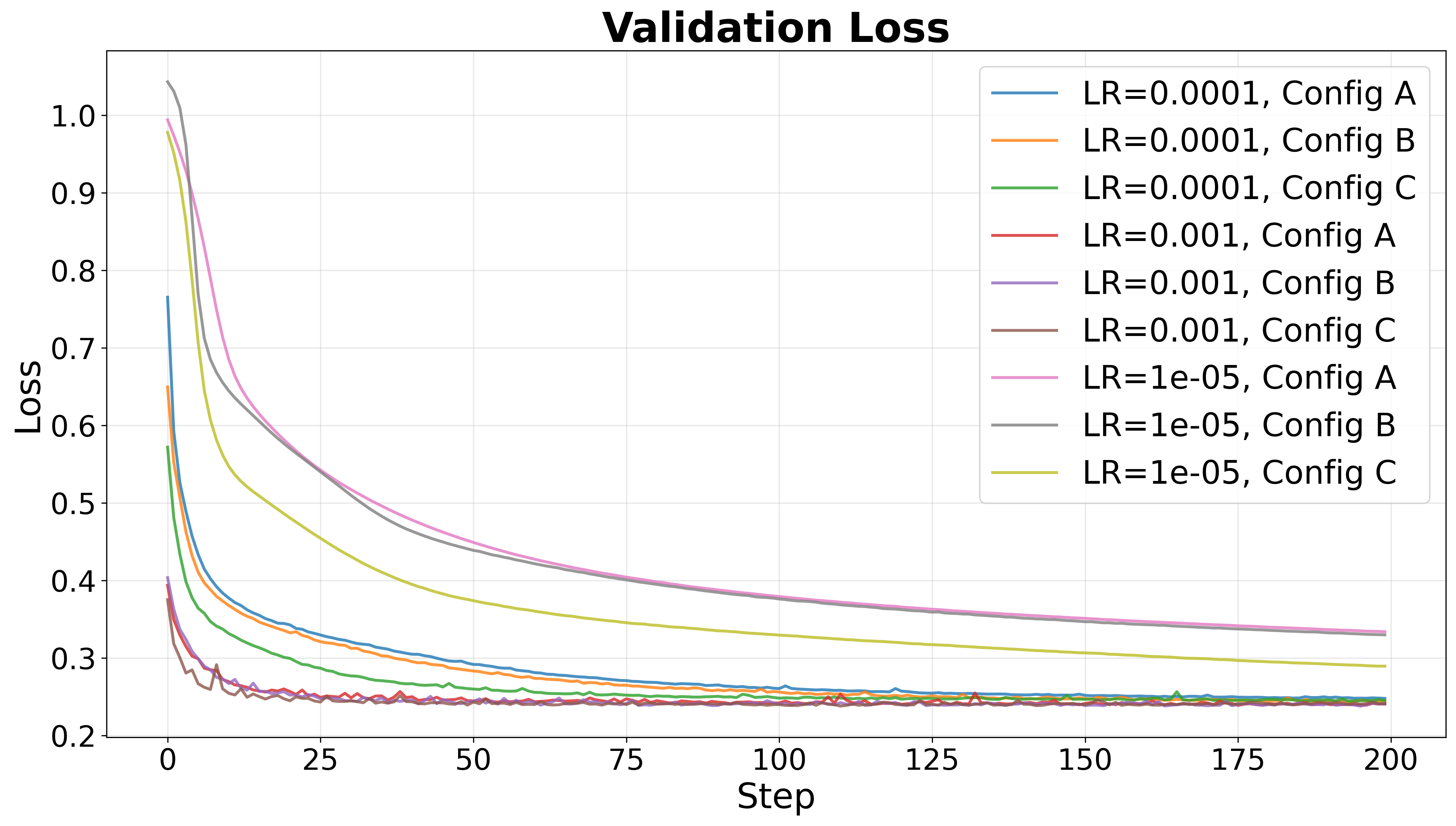}
    \end{minipage}
    \caption{Training loss (left) and validation loss
    (right) as a function of epoch. The decreasing trends indicate convergence and
    generalization, while the non-zero floor reflects irreducible counting noise
    in the counts channel and non-stationary observing conditions encoded in the
    exposure channel.}
    \label{fig:learning_curves}
\end{figure}

Figure~\ref{fig:learning_curves} shows the learning curves for the selected model.
The training loss decreases steadily as a function of epoch, indicating that the
ConvLSTM converges to a stable solution and is able to extract non-trivial temporal
regularities from the input sequences. The validation loss follows a broadly
consistent trend, supporting the interpretation that the learned dynamics generalize
to held-out windows rather than reflecting pure memorization of the training set.
At the best validation epoch, the final losses reach
$\mathrm{MSE}_{\mathrm{train}} \simeq 0.265$ and
$\mathrm{MSE}_{\mathrm{val}} \simeq 0.238$
(values reported for the best validation epoch).

The model output can be interpreted as a \emph{predictive baseline} for the nominal
evolution of the daily sky maps: for the bulk of windows---i.e., away from abrupt changes in observing conditions---the prediction captures the large-scale morphology and the smooth day-to-day variability, and the residual
field is dominated by low-amplitude fluctuations.

The learning curves do not indicate perfect prediction, and a structured residual
component persists. As already noted from the best-epoch training and validation
losses discussed above, both curves approach a non-zero plateau. The fact that the
validation loss remains of the same order as the training loss, rather than diverging
from it, suggests that the residual floor is not primarily a consequence of
overfitting, but reflects limitations intrinsic to the data and to the predictive
task itself.
This behavior is expected given both (i) the intrinsic
stochasticity of the photon counts and (ii) the heterogeneous nature of the two
input channels. Count maps are subject to Poisson fluctuations whose variance depends
on local intensity and exposure, implying an irreducible prediction error even under
a perfectly specified generative model. In addition, our pipeline combines simulated
photon counts (\texttt{gtobssim}) with exposure maps derived from real spacecraft
pointing and livetime. While this choice increases realism, it also introduces
non-stationarities unrelated to astrophysical variability: pointed observations,
repointings, departures from standard survey mode, and operational constraints can
induce sharp day-to-day exposure modulations that locally break the smooth temporal
correlations that recurrent models such as ConvLSTMs are biased to exploit.

This has direct implications for anomaly detection. Without calibration, exposure-driven
prediction errors could be misinterpreted as candidate transients. For this reason,
the learning curves must be interpreted together with the downstream statistical
treatment of residual maps. In our framework, detection thresholds are derived
empirically from training residuals on a per-pixel basis
(Section~\ref{sec:threshold}), so that regions that are systematically harder to
predict naturally receive higher thresholds. Finally, spatial coherence is enforced
through local-density or connected-component filtering, suppressing isolated
supra-threshold pixels while retaining compact residual ``blobs'' consistent with
localized transient emission.

In the next subsection, we quantify how these calibrated residual maps translate into
frame-level and pixel-level detections, and we illustrate representative examples of
localized anomalies recovered by the full pipeline.

\subsection{Characterization of the source population from the anomaly detection procedure}
To test and validate the proposed procedure, and to assess the performance of the ConvLSTM architecture, we cross-correlated the anomaly-detection source list obtained from the first 16 days of operation with the 4FGL$-$DR4 general source catalog \citep{Ballet2023_4FGLDR4}. We then compared variability-related properties of the cross-matched sample with those of the full catalog. We adopted cross-correlation radii of $1^{\circ}$ and $2^{\circ}$, since the localization accuracy of the anomaly-detection output is relatively poor. As a consequence, unique associations cannot generally be established: for each detected anomaly, multiple 4FGL$-$DR4 sources may fall within the matching radius. Nevertheless, the ensemble of associated sources can be compared statistically with the full catalog.
\begin{figure}[h!!]
    \centering
    \includegraphics[width=0.8\linewidth]{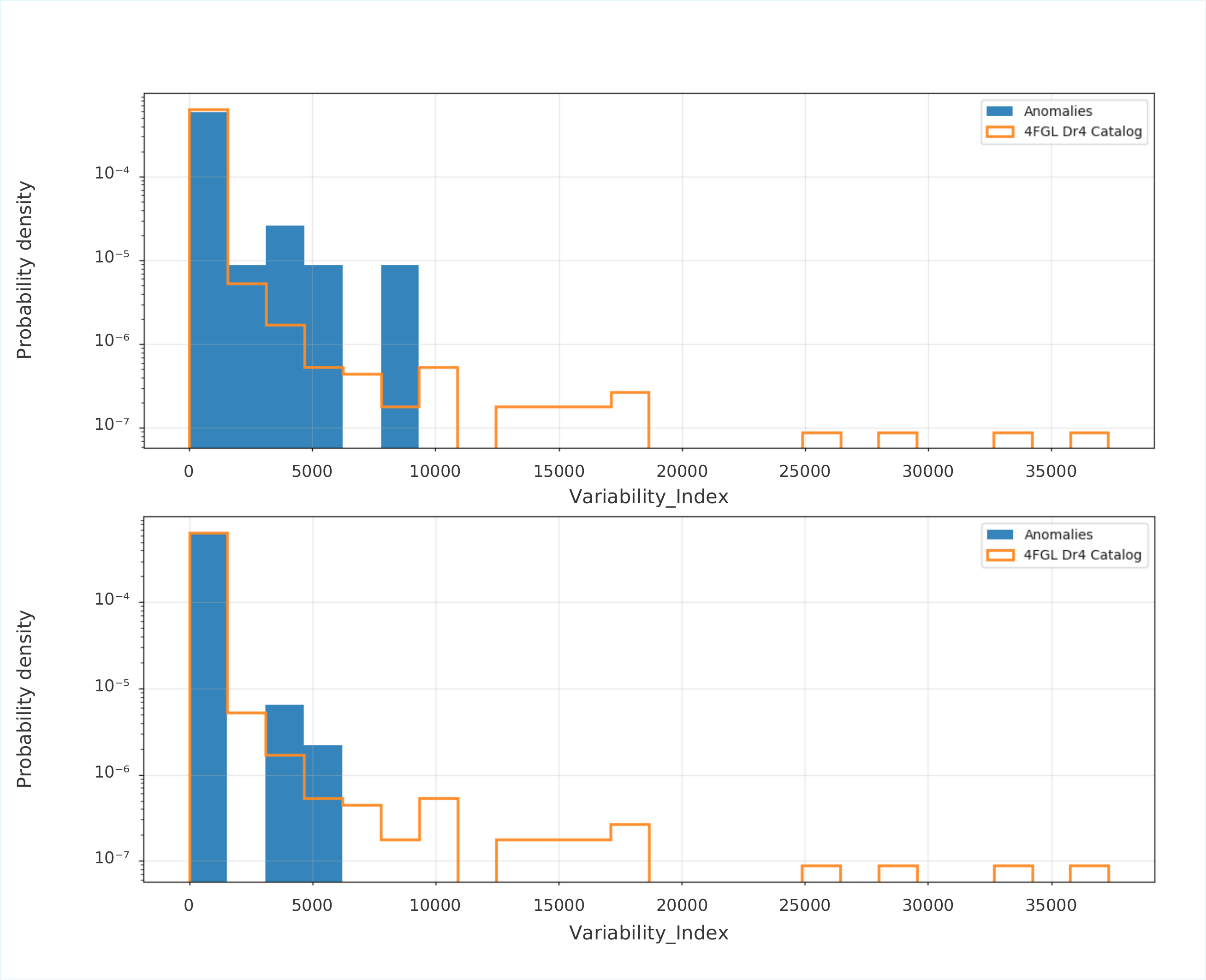}
    \caption{Normalized histograms of Variability\_Index variable for sources associated to the output of anomalies showed in step-filled blue superimposed with all 4FGL$-$DR4 catalog sources showed in orange. Top panel: cross correlation radius of 2 degrees; bottom panel: cross correlation radius of 1 degrees}
    \label{fig:hist}
\end{figure}

We investigated differences in terms of the variability index \citep{Ballet2023_4FGLDR4}, which quantifies the degree of flux variability of a source as a function of time. The resulting distributions are significantly different: a Kolmogorov--Smirnov test yields a $p$-value below 0.05 in both cases, indicating that the two samples are unlikely to originate from the same parent population. This result suggests that the anomaly-detection procedure preferentially selects variable sources, as expected. The two populations are shown in Figure~\ref{fig:hist}.

Overall, this provides a validation of the method, demonstrating its ability to register flux variations as anomalies with respect to the averaged reconstructed skies.

\subsection{Discussion of two particular cases}

\begin{figure}[h!!]
    \centering
\includegraphics[width=0.8\linewidth, angle=270]{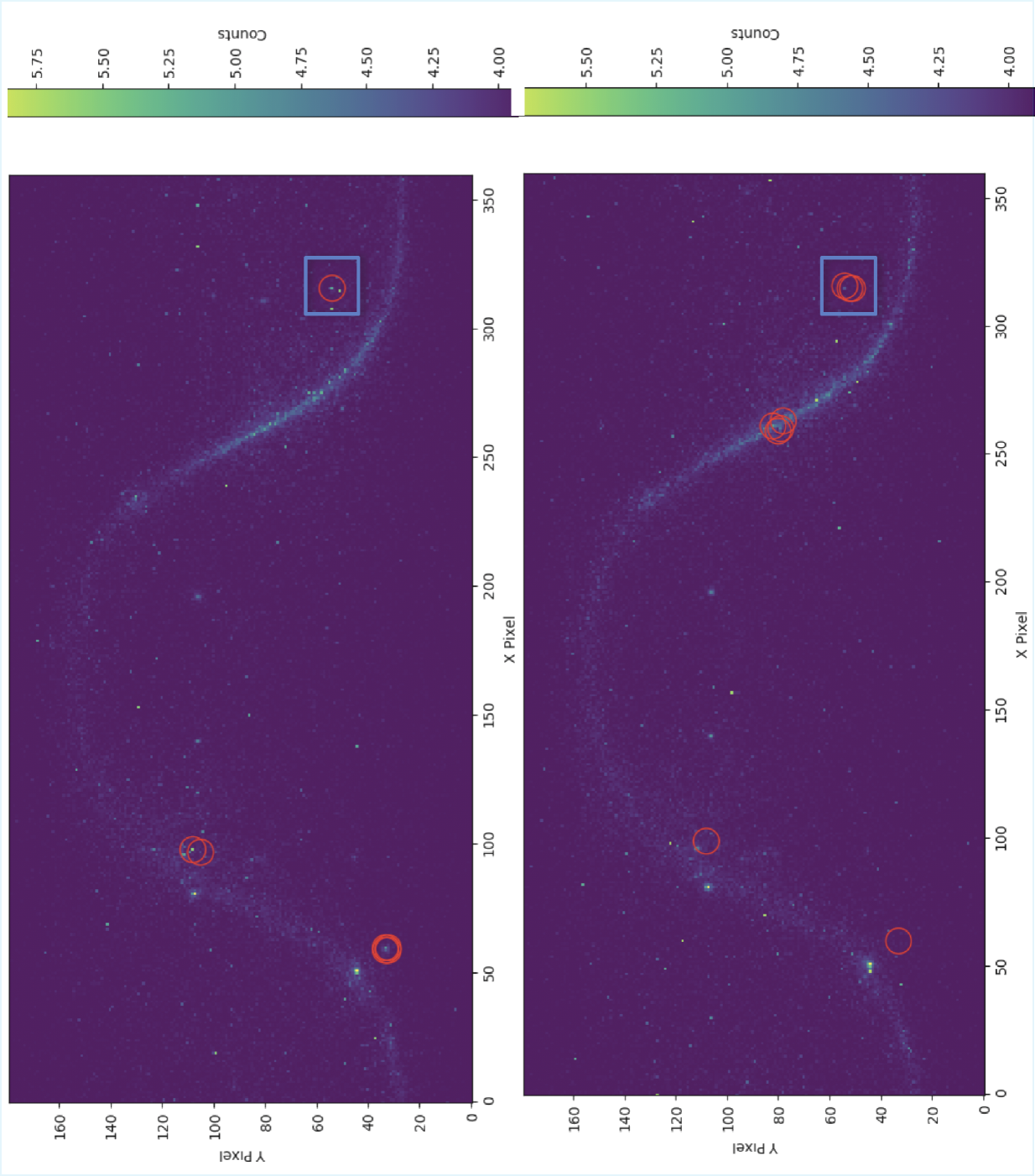}
    \caption{True-sky maps showing the list of anomalies indicated by red circles, while the GRB localization is highlighted by a blue square superimposed on them. Top panel: 2008-09-16 time frame; bottom panel: 2008-09-17 time frame.}
    \label{fig:GRB}
\end{figure}

We tested our procedure on a few representative cases, such as transient sources (e.g., GRBs) and exceptionally bright flaring activity from well-known sources (e.g., blazars).
We then examined the anomaly maps produced by the ConvLSTM pipeline over time windows containing known transient events.
We use the term \emph{detection} in an operational sense consistent with the anomaly criterion defined in Section~\ref{sec:threshold}. In particular, we detected GRB~080916A on the day of the GRB prompt emission and Fermi-LAT detection \citep{gcn8263} (Mission Elapsed Time interval from 243216766~s to 243303168~s), as well as a \emph{post-transient over-prediction} on the following day, i.e., a lagged predictive artifact in which the model continues to predict an excess after the transient has disappeared. Figure~\ref{fig:GRB} shows the true-sky maps with the detected anomalies highlighted as red circles.

\begin{figure}[h!!!!]
    \centering
\includegraphics[width=0.8\linewidth]{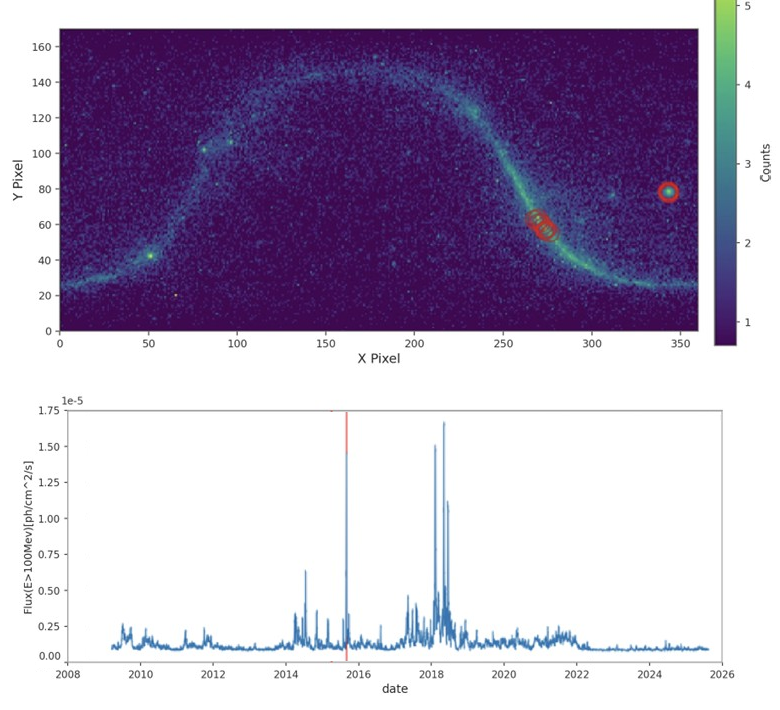}
    \caption{Top panel: True sky with indicated the list of anomalies as red circles with the detection of an anomaly associated to 3C 279 (red circle above the galactic plane). Lower panel:  3C 279 weekly time bin light curve, the vertical red line indicates the time frame relative to the sky analyzed in the top panel.}
    \label{fig:3C279}
\end{figure}

Furthermore, our procedure is able to detect strong flaring activity from well-known sources such as blazars. In particular, we detected a strong outburst from the FSRQ~3C~279 on 2015-06-14, as shown in Figure~\ref{fig:3C279} (Mission Elapsed Time interval from 455934465~s to 456193665~s). This flare has been reported by both Fermi-LAT and AGILE in gamma rays, as well as at other wavelengths (see \citealt{atel7633}, \citealt{atel7631}, and \citealt{atel7639}; see also \citealt{ackermann2016minute}).

\section{Conclusions and Future Work}
\label{sec:conclusion}

In this work we presented an end-to-end, fully reproducible framework to study transient detection in a wide–field-of-view gamma-ray monitoring scenario, by combining long-duration sky simulations with self-supervised spatio-temporal deep learning. Starting from a ten-year \texttt{gtobssim} simulation, we constructed daily all-sky maps of counts and exposure and used them as a two-channel time series of images. We trained a ConvLSTM-based model to learn the nominal evolution of the sky and to provide next-frame predictions (equivalently, reconstructions in a predictive setting). Departures from this learned baseline were quantified through pixel-wise mean-squared residual maps, which were then converted into anomaly candidates through statistically motivated thresholds and spatial-coherence filtering.

We found that the ConvLSTM is able to learn non-trivial temporal regularities in the simulated dataset, yielding stable learning curves and residual maps whose structure can be calibrated empirically from the training distribution. The downstream anomaly-detection stage produces candidates that are consistent with variability-driven phenomena: when cross-correlating the first days of detections with the 4FGL--DR4 catalog, the associated population shows a statistically significant shift toward higher variability indices with respect to the full catalog, supporting the interpretation that the procedure preferentially highlights variable sources. In addition, targeted checks on specific events demonstrate that the pipeline can flag both fast transients (e.g., GRB~080916A) and strong flaring activity from bright blazars (e.g., the 2015 outburst of 3C~279), illustrating the practical sensitivity of the residual-based approach to localized, time-dependent excesses.

A key methodological takeaway is that realistic observing conditions matter: using real spacecraft-derived exposure increases fidelity but also introduces non-stationarities that can limit purely autoregressive predictability at daily cadence. Within our framework, this motivates (and partially justifies) the combination of a learned predictive baseline with adaptive, data-driven residual calibration and spatial post-processing, rather than relying on raw reconstruction errors alone.

Several extensions naturally follow from this first study. A first, straightforward improvement is to move beyond the energy-integrated representation adopted here and train the ConvLSTM on energy-resolved sequences, so as to leverage the spectral signatures of transients and mitigate confusion between genuine astrophysical variability and background- or exposure-driven effects.

In the same spirit, the predictive model could be explicitly conditioned on observing-state information—e.g., survey-mode indicators, livetime, rocking angle, or compact summaries of the recent pointing history—to better disentangle operational discontinuities from genuine sky variability. Additional improvements are likely to come from increasing the spatial and temporal adaptivity of the analysis pipeline. On the spatial side, adopting a finer pixelization (where computationally feasible) would reduce source confusion and localize excesses more precisely, facilitating the association of a detected gamma-ray transient with plausible multiwavelength counterparts. On the temporal side, enabling flexible (and potentially adaptive) time binning would make the framework effective across a wide dynamic range of transient timescales, from fast events such as GRBs to slower phenomena evolving over days to months. Further gains are also expected from a more principled treatment of uncertainty: replacing deterministic predictions with probabilistic outputs, or explicitly modeling count statistics conditional on exposure, would yield better-calibrated anomaly scores and decision thresholds with a clearer statistical interpretation.

A complementary direction is systematic benchmarking through controlled injections of synthetic transients with known properties into the simulated dataset, enabling quantitative measurements of detection efficiency and false-positive rates as functions of flux, duration, and sky location, and allowing direct comparison with alternative architectures such as convolutional autoencoders or attention-based sequence models. Finally, once thoroughly validated in simulation, the same pipeline can be transferred to Fermi-LAT daily maps to assess robustness under fully realistic conditions and to explore applications closer to near-real-time transient alerting.

Overall, the proposed simulation-to-inference chain provides a controlled testbed for developing and evaluating anomaly-detection strategies on long-duration datasets, while remaining close enough to real observing conditions to expose the main practical challenges faced by automated transient searches.

\vspace{6pt} 
\section*{CRediT authorship contribution statement}
Conceptualization: A.G., S. S., A.V., S.C., M.M. (Marcello Marconi), F.L. and E.P.; methodology, A.G, S.S., A.V., A.M., S.C., M.M. (Matteo Martini), F.F., R.G., E.W.D.L., U.D.M. and S.M.; software, S.S., A.V. and A.M.; validation, S.S., A.V., S.C., M.M. (Matteo Martini), F.F., R.G., E.W.D.L., U.D.M. and S.M.; writing---original draft preparation, S.S., S.C.; writing---review and editing, S.S.; project administration, A.G., M.M. (Marcello Marconi), E.P., S.C., F.F. and R.G.; funding acquisition, A.G., M.M. (Marcello
Marconi) and E.P. All authors have read and agreed to the published version of the manuscript.

\section*{Funding}
This paper is supported by the Fondazione ICSC, Spoke 3 Astrophysics and Cosmos Observations. National Recovery and Resilience Plan (Piano Nazionale di Ripresa e Resilienza, PNRR) Project ID CN00000013 ``Italian Research Center on High-Performance Computing, Big Data and Quantum Computing'' funded by MUR Missione 4 Componente 2 Investimento 1.4: Potenziamento strutture di ricerca e creazione di ``campioni nazionali di R\&S (M4C2-19)'' - Next Generation EU (NGEU)

\section*{Data Availability}
The simulated dataset generated in this study is publicly available on Hugging Face at \url{https://huggingface.co/datasets/Idea-re/fermi-lat-synthetic-daily-sky-maps}. The public release includes the original energy-resolved tensor used in this work, with shape $(3649, 2, 10, 360, 180)$, split into four NumPy files to facilitate download and reuse. The repository also provides accompanying documentation and metadata, including a README file, a machine-readable metadata description, and an example Python script for loading the data.

\section*{Declaration of competing interests}
The authors declare that they have no known competing financial interests or personal relationships that could have appeared to influence the work reported in this paper.

\section*{Declaration of Generative AI and AI-assisted technologies in the writing process}

During the preparation of this work the authors used ChatGPT to improve language, style, and readability. The scientific content, structure, and conclusions were determined by the authors. After using this tool, the authors reviewed and edited the manuscript as needed and take full responsibility for the content of the publication.

\bibliographystyle{elsarticle-harv} 
\bibliography{references}

\end{document}